\documentclass[11pt,a4paper]{article}
\usepackage[utf8x]{inputenc}
\usepackage{ucs}
\usepackage[left=2.5cm,top=3cm,right=2cm,bottom=3cm]{geometry}
\usepackage{authblk}

\usepackage{cite}

\usepackage{amsmath,amsfonts,bm,amssymb,latexsym,amsthm}
\usepackage{graphicx, graphics}
\usepackage{multirow}
\usepackage{microtype}
\usepackage[none]{hyphenat}
\usepackage{setspace} 
\usepackage{array}
\usepackage{amsfonts}   
\usepackage[bottom]{footmisc}
\usepackage{color}
\usepackage{xurl}
\usepackage{float}
\usepackage{rotating}

\usepackage{soul} %Flix added, for highlights
\usepackage{todonotes} %Flix added, for notes

\usepackage{placeins}
\usepackage{multicol}
\usepackage{multirow}

\usepackage{morefloats}
\usepackage{booktabs}
\usepackage{multirow}
\usepackage{pifont}

\usepackage{svg}

\usepackage{comment}

\newcolumntype{C}[1]{>{\centering\let\newline\\\arraybackslash\hspace{0pt}}m{#1}}

\usepackage{hyperref}
\hypersetup{colorlinks, breaklinks, linkcolor=[rgb]{0,0.3,1}, citecolor=[rgb]{0,0.3,1}, urlcolor=[rgb]{0,0.3,1} }
\usepackage{longtable}

\usepackage{lmodern} %% added to compile in bash

\begin{document}

%------------------------------------------
%\title{Regional emission dynamics across EU ETS phases}
\title{Regional emission dynamics across phases of the EU ETS}

\author[1,*]{Marco Due\~nas}
\author[1,2]{Antoine Mandel}

\affil[1]{\small CLIMAFIN -- Climate Finance Alpha, Paris, France}
\affil[2]{\small Centre d'Economie de la Sorbonne -  Paris School of Economics- CNRS-Universit\'{e} Paris 1 Panth\'{e}on-Sorbonne, Paris, France}
\affil[*]{Corresponding author email: marco.duenas@climafin.com}

\date{}

\maketitle
\vspace{-40pt}

\begin{abstract}
This paper explores the relationship between economic growth and CO$_2$ emissions across European regions from 1990 to 2022, specifically concerning the dynamics of emissions growth rates through different phases of the European Union Emissions Trading System (EU ETS). We find that emissions dynamics exhibit significant volatility influenced by changing policy frameworks. Furthermore, the distribution of emissions growth rates is asymmetric and displays fat tails, suggesting the potential for extreme emissions events. We identify marked disparities across regions: less developed regions experience higher emissions growth rates and greater volatility compared to many developed areas, which show a trend of declining emissions and reduced volatility. Our findings highlight the sensitivity of emissions to policy changes and emphasise the need for clear and effective governance in emissions trading schemes.
\end{abstract}
\bigskip
\noindent \textbf{Keywords:} EU ETS; CO$_2$ emissions; Growth volatility: Scaling effects; Regional development

\flushbottom
\maketitle
\thispagestyle{empty}

%\noindent Please note: Abbreviations should be introduced at the first mention in the main text – no abbreviations lists. Suggested structure of main text (not enforced) is provided below.

%Possible journals:
%https://iopscience.iop.org/journal/1748-9326
%https://www.sciencedirect.com/journal/advances-in-climate-change-research
%https://agupubs.onlinelibrary.wiley.com/journal/23284277
%https://www.scimagojr.com/journalsearch.php?q=21101065239&tip=sid&clean=0

\section*{Introduction}

%TODOAM:EU-ETS reform, an efficient climate policy implies a clear trend in emission reduction, ideally less volatile than GDP. Also the issue of phases and of the EU-ETS reform more clearly related to results.
%What would be an objective in terms of emissions growth rates and emissions. Rather than decoupling, it seems that emissions amplify macro fluctuations. 

Since the 1990s, numerous international efforts have shown commitment to reducing emissions, with the European Union (EU) often leading these initiatives and gradually implementing modifications to improve a cap-and-trade system \cite{sato2022allocation}. The adoption of the United Nations Framework Convention on Climate Change (UNFCCC) in 1992 marked the beginning of global efforts, which were further strengthened by the Kyoto Protocol in 1997, the launch of the EU ETS in 2005, the Copenhagen Accord in 2009 and the Paris Agreement in 2015.

In the evolving landscape of global climate policy, the EU ETS stands as a significant endeavour designed to reduce greenhouse gas (GHG) emissions cost-effectively by setting a cap on emissions and enabling trading among emitters. Established in 2005, the EU ETS is recognised as one of the world's major carbon markets. As the impacts of climate change become increasingly critical, the effectiveness of such market-based mechanisms in achieving real emissions reduction and their economic impact remains a crucial area of research \cite{goulder2013carbon, ellerman2016european, dechezlepretre2023joint}. 

This paper provides a comprehensive analysis of the changes in emissions volatility across different phases of the EU ETS phases, empirically examining the dynamic relationship between economic growth and CO$_2$ emissions growth rate volatility in European regions. It highlights significant patterns in how emissions volatility correlates with economic fluctuations and assesses the impact of policy changes on this relationship. Our findings support the argument that an efficient climate policy should aim for a stable emission reduction trend, ideally less volatile than GDP fluctuations. However, despite the necessity of such stability for long-term planning and the adoption of green technologies, our empirical findings show that emissions continue to fluctuate significantly with the business cycle, posing challenges to decoupling emissions from economic activity.

Climate change is primarily driven by pollutant emissions, which are influenced by productivity shocks, economic crises, and technological changes. Consequently, understanding how GHG emissions dynamics are intertwined with economic activity can contribute to a discussion on sustainable development to facilitate the achievement of emissions reductions. The economic cycle plays a key role in the effectiveness and efficiency of emissions trading schemes. During periods of economic growth, increased production and consumption typically lead to higher emissions, thereby boosting the demand for emission permits and potentially increasing their prices. Conversely, during recessions, a decline in economic activity results in lower emissions and reduced demand for emission permits, which decreases their prices. These fluctuations can significantly impact investment incentives in clean technologies: higher carbon prices may encourage investments, while lower prices could discourage them. Thus, economic volatility is expected to introduce significant fluctuations in carbon prices, posing challenges to the market's stability and predictability, which are crucial for achieving long-term emissions reduction.

Since the objective of decoupling emissions from economic activity faces significant challenges due to the robust and persistent correlation between emissions and economic growth \cite{knight2014economic}, a critical debate has emerged about whether economic growth can coexist with the urgent need to mitigate emissions. This discussion has focused on whether wealthy countries should adopt a steady-state or degrowth strategy while also exploring alternative economic models such as the circular economy and green technologies \cite{knight2014economic, ghisellini2016review, murray2017circular}. Additionally, there is growing interest in the role of international cooperation in harmonising environmental policies with economic development \cite{wto2023reglobalization}, indicating that such policies are increasingly impacting emission trends, as reflected in the economic practices within countries, contributing to our understanding of the decoupling mechanism \cite{haberl2020systematic}.

The empirical evidence demonstrates that GHG emissions and their pricing are closely synchronised with the business cycle \cite{peters2012rapid, fischer2013environmental, doda2014evidence}. The correlation between fossil fuel carbon intensity (FFCI) and CO$_2$ emissions suggests that during economic downturns, countries typically engage in less energy-intensive activities, therefore, reducing emissions. However, the 2008-2009 global financial crisis presented a contrasting outcome, with emissions not decreasing as expected because the use of fossil fuels was promoted to overcome the crisis \cite{peters2012rapid}. Further analysis of emissions and GDP decomposition into their trend and cyclical components shows that CO$_2$ emissions generally increase during periods of economic expansion, particularly in higher-income countries, highlighting their procyclicality \cite{doda2014evidence}. Additionally, although emissions tend to be more volatile than GDP, this volatility decreases as GDP per capita increases \cite{doda2014evidence}.

Additionally, emissions respond asymmetrically to the business cycle, decreasing more significantly during economic contractions than they increase during expansions \cite{sheldon2017asymmetric}. This asymmetry suggests that during downturns, emissions reduce more than production levels, likely because firms prioritise energy-efficient operations. Conversely, during economic upturns, firms may invest more heavily in new technologies, including environmentally friendly ones, potentially reducing the intensity of emissions even as they increase energy-intensive operations. %%However, the limited ability to invest in new technologies during downturns may result in less significant reductions in emissions intensity, leading to higher emissions than might be expected given the scale of economic contraction.

Recent studies have explored the impacts of the EU ETS, revealing interdependencies between policy-induced economic incentives and environmental outcomes \cite{koch2014causes, bayer2020european, sato2022allocation}. Despite attempts by European countries to decouple emissions from economic activity, the extent of this decoupling has varied significantly across countries, sectors, and pollutants \cite{naqvi2017fifty, ivanova2017mapping}. Moreover, the trend towards decoupling has not been consistent and is vulnerable to fluctuations in environmental policies and economic crises, resulting in wide disparities among regions \cite{peters2012rapid, bayer2020european}. This suggests a diverse array of decoupling trajectories.

The impact of regional disparities and economic activities on emissions, particularly the volatility of emissions growth rates across different phases of the EU ETS, has received less attention. Understanding these dynamics is crucial as it can help refine current and future environmental policies.

This study builds on existing literature by examining the relationship between economic growth and CO$_2$ emissions growth rates across European regions from 1990 to 2022. This period encompasses both the pre- and post-implementation phases of the EU ETS. Our findings indicate that the distribution of emissions growth rates exhibits significant asymmetry and fat tails. The latter suggests a higher likelihood of extreme emissions changes. Additionally, we discovered a robust negative correlation between the volatility of emissions growth rates and the economic size of regions, indicating that smaller regions (using their GDP as a proxy) experience higher emission volatility. 

We have implemented an econometric model to analyse the dynamics of emissions growth rates and their relationships with emission levels, GDP shocks, and the volatility scale relationship with regional economic size within a framework that accommodates the evidence of fat tails. The analysis, which uses moving windows for different periods, reveals varying impacts on emission dynamics that can be primarily attributed to the policy changes across different phases of the EU ETS.

We find that over time, the impact of economic shocks on emissions has intensified following the implementation of the EU ETS. Although there is a general downward trend in regional emissions, our estimates indicate considerable uncertainty in the emissions rates and instability in their fat-tailed distribution, which also shows varying levels of asymmetry. These findings are crucial for policymakers, as the persistence of these distributional characteristics through different phases of the EU ETS underscores the challenges in achieving stability and guiding regions towards more defined paths for decarbonisation.

\section*{Results}

\subsection*{The distribution of emissions growth rates}

Since the 1990s, European regions have achieved some reductions in their GHG emissions \cite{EDGAR_report}, making progress towards decoupling economic growth from emissions. This decoupling alludes to the fact that while economies continue to grow, emissions are not increasing at the same rate, although the average emissions growth rates are rather heterogeneous when looking at regions with different development levels (see Supplementary Information). 

However, although most regions show a trend moving towards reducing their emissions, driven by the demands of sustainable growth to fight climate change, the processes by which they achieve this are of significant interest. Regions can exhibit smooth or highly erratic trends in their emissions reductions. This variability links to how businesses within these regions adapt to regulations to reduce their emissions intensity. Consequently, understanding how emissions shocks manifest and their connection to GDP shocks becomes critically important.

\begin{figure}[h!]
\centering
\includegraphics[width=\linewidth]{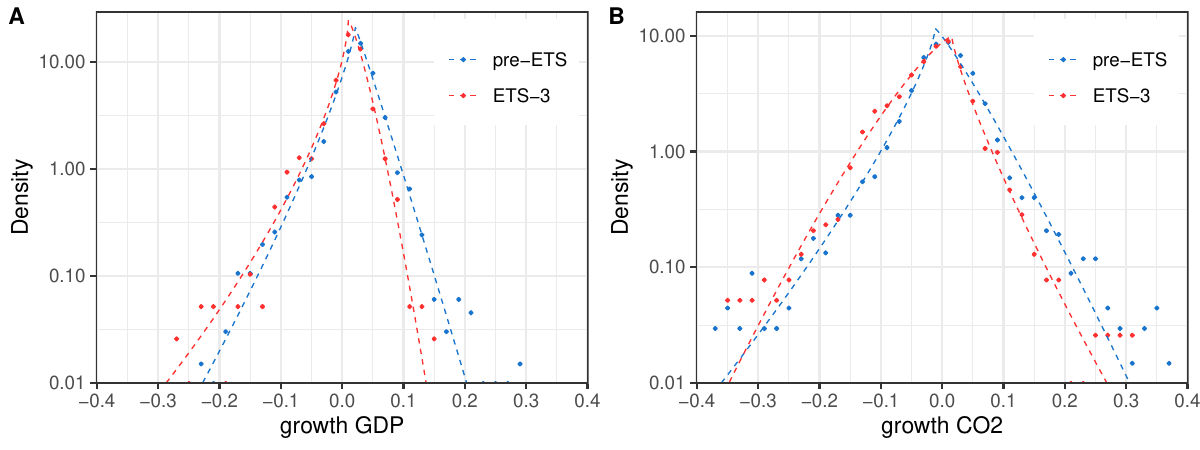}
\caption{Probability density functions estimations of GDP and emissions growth rates for the pre-EU ETS (1990-2004) and EU ETS-3 (2013-2020) periods. Panel A corresponds to GDP growth rates. Panel B corresponds to CO$_2$ emissions growth rates. Dashed lines represent the parametric estimation of the corresponding AEP distributions; see Eq.~\eqref{eq:subb}.}
\label{fig:pdf_emisions}
\end{figure}

Figure~\ref{fig:pdf_emisions} shows the empirical distributions of GDP and CO$_2$ emissions growth rates for the pre-EU ETS and ETS-3 periods. The figures also include the corresponding parametric maximum likelihood estimation of the probability density functions.
%----------------------------------------
%----------------------------------------

Growth rates for both GDP and CO$_2$ emissions exhibit fat-tailed distributions. The term fat-tailed refers to the phenomenon where the tails of the distribution decrease much more slowly than would be observed if the growth rates were dictated by a normal distribution -- if that were the case, the distribution would have a parabolic shape (note that the probability density axis is on a logarithmic scale). Conversely, the tails exhibit a tent-shaped form. This indicates an excess of kurtosis compared to the normal distribution, meaning there is a greater concentration of probability mass around the distribution mode and, simultaneously, a higher probability of observing either positive or negative extreme events. 

While this feature is relatively well-known for GDP growth rates \cite{lee1998universal, castaldi2009patterns, campi2020volatility}, it is interesting to find that CO$_2$ emissions growth rates exhibit similar behaviour. Beyond the clear link between GHG emissions and economic activity, the shape of the distribution constitutes evidence of the complex dynamics that govern emissions.
 
We use a parametric approach using the asymmetric exponential power distribution (AEP) to determine the distribution of growth rates. When comparing the periods before the EU ETS implementation and during its third phase (ETS-3), we observed notable changes in the distributions of GDP and CO$_2$ emissions growth rates, characterised by persistent asymmetry. Further analysis of the tails' shapes and their distances revealed these changes clearly. 

We used the Laplace distribution as a reference point, which is well-known for fitting GDP growth rate distributions \cite{lee1998universal, castaldi2009patterns}. This distribution corresponds to AEP shape parameters $b_l = b_r = 1$ and scale parameters $a_l = a_r$, as specified in Eq.~\eqref{eq:subb} in the methods section. From the shape parameters $b$, we can infer the frequency of extreme events -- a value less than one indicates a higher propensity for such events. The scale parameters $a_l$ and $a_r$ inform us about the distribution's breadth, directly relating to its dispersion and volatility. It is crucial to note that the fat-tail behaviour suggests that both shape and scale parameters influence the variance of the distribution.

The estimates of the distribution parameters, reported in Table~\ref{tb:pdfs}, show that for the GDP growth rates, the scale parameters slightly decreased during the ETS-3 period compared to pre-ETS period, the left tail shape parameter decreased, and the corresponding right tail parameter increased. This pattern suggests a reduction in overall volatility, with a higher likelihood of extreme negative events and a lower likelihood of extreme positive events occurring. Furthermore, the distribution mode is positive in both periods, though it is smaller in the latter period. Visually, these changes are observed as an apparent clockwise rotation of the distribution around its peak (as seen in Fig.~\ref{fig:pdf_emisions}A).
%*-*-*-*-*-*-*-*-*-*-*-*-*-*-
\begin{table}[h!]
\begin{center}
\scriptsize 
\caption{Estimated AEP distributions of GDP and CO$_2$ emissions growth rates corresponding to the pre-EU ETS (1990-2004) and EU ETS-3 periods (2013-2020).}\label{tb:pdfs}\vspace{3pt}
\resizebox{0.65\textwidth}{!}{\renewcommand{\arraystretch}{1.1}
\begin{tabular}{l cc cc } 
\toprule 
 & \multicolumn{2}{c}{GDP growth rates} & \multicolumn{2}{c}{CO$_2$ growth rates} \\ 
\cmidrule(lr){2-3}
\cmidrule(lr){4-5}
 & pre-ETS & ETS-3 & pre-ETS & ETS-3 \\
\midrule
$b_l$ & 0.804(0.028) & 0.657(0.026) & 0.778(0.026) & 1.192(0.059) \\
$b_r$ & 1.047(0.043) & 1.288(0.066) & 1.135(0.039) & 0.876(0.046) \\
$a_l$ & 0.026(0.001) & 0.025(0.001) & 0.039(0.001) & 0.063(0.002) \\
$a_r$ & 0.025(0.001) & 0.021(0.001) & 0.051(0.001) & 0.032(0.001) \\
$m$ & 0.021(0.001) & 0.010(0.000) & -0.011(0.001) & 0.011(0.001) \\
\bottomrule
\multicolumn{5}{p{8cm}}{\textit{Notes:} Standard errors are in parentheses.}\\
\end{tabular}}
\end{center}
\end{table}
%*-*-*-*-*-*-*-*-*-*-*-*-*-*-

A similar analysis of the CO$_2$ emissions growth rates also reveals changes in the shape and scale parameters. To start, compared to GDP growth rates, the wider stretched tails of emissions confirm that emissions are much more volatile \cite{doda2014evidence}. For the CO$_2$ emissions growth rate distributions, we note an increase in the scale parameter for the left tail and a decrease for the right tail. Simultaneously, the shape parameter for the left tail increases, while it decreases for the right tail, and a shift of the distribution mode from negative to positive. Although changes in the volatility pattern are less evident than those observed for GDP growth rates, these adjustments suggest a higher number of observations in the left tail, indicative of thinning and increased variance in this tail. Despite this, there remains a relatively higher propensity to observe extreme positive emission events.

In conclusion, while both GDP and CO$_2$ emissions growth rates undergo transformations in their distribution shapes, CO$_2$ emissions display significantly greater volatility compared to GDP. This greater fluctuation in emissions than in economic growth may reflect the varying impacts of policy measures, economic conditions, or external environmental factors. Notably, the increase in the dispersion parameter for the left tail of CO$_2$ emissions and a shift towards more fat-tailed behaviour in the right tail suggest a decreased likelihood of extreme emission reductions but an increased likelihood of significant emission surges, which to a great extent agrees with the observed pattern of emission trend reductions of most European regions \cite{EDGAR_report}. However, this pattern also means that despite a general trend towards emissions reductions, there are substantial spikes in emissions. These spikes could result from inconsistent policy implementation and the unique nature of economic activities that directly influence emissions.

\begin{figure}[h!]
\centering
\includegraphics[width=\linewidth]{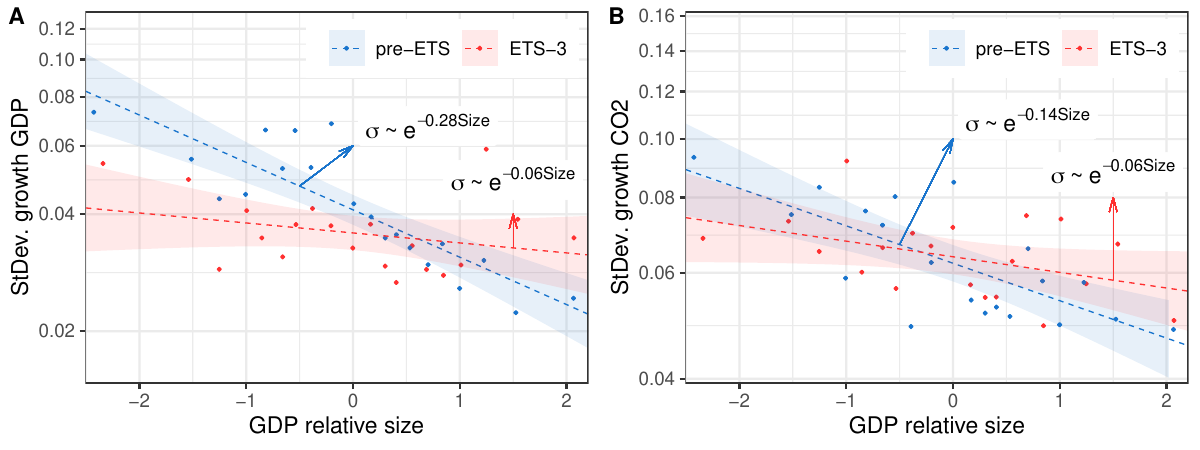}
\caption{Scaling relationship between growth rates volatility and economic size using binned statistics. Panel A corresponds to GDP growth rates. Panel B corresponds to CO$_2$ emissions growth rates. Dashed lines depict the linear trend estimations, including error bands.}
\label{fig:sd_size}
\end{figure}

Next, we analyse whether the emissions and GDP growth rates volatility are distributed uniformly across European regions. In the case of GDP growth volatility, the literature recognises a marked negative relationship between the standard deviation of GDP growth rates and the size of the economies, indicating that volatility diminishes as size increases following a power-law distribution \cite{lee1998universal, castaldi2009patterns, campi2020volatility}. This scaling relationship is also observed for GDP per capita growth rates and firms' sales growth rates \cite{stanley1996scaling, sutton2002variance, bottazzi2006explaining, riccaboni2008size}. In Fig.~\ref{fig:sd_size}A, we used binned statistics to estimate this scaling parameter, finding values of $\beta=-0.28$ for the pre-ETS period and $\beta=-0.06$ for the ETS-3 period.  
%----------------------------------------
%----------------------------------------

%Before presenting the results related to CO$_2$ emissions growth rates, it is necessary to mention a remarkable conclusion drawn from this statistical regularity of GDP growth volatility.

A remarkable conclusion drawn from this growth pattern is that most economic systems must be seen as complex organisations composed of interacting units, not just as collections of independent entities \cite{lee1998universal, stanley1996scaling, sutton2002variance, bottazzi2006explaining, riccaboni2008size, castaldi2009patterns, campi2020volatility}. This idea is based on the fact that if we envision an economic system as a collection of independently and identical growing units -- as a consequence of the law of large numbers, the variance of the sample means decreases as the sample size increases -- the scale parameter would be $\beta=-0.5$. On the other hand, in a scenario where units are highly interdependent, a shock involving a given entity will propagate to the other units and, therefore, the growth volatility of one unit will affect others. In this scenario, the volatility across all units would be identical, and therefore, the scale parameter would be $\beta=0$. Hence, the estimated coefficients between -0.5 and 0 suggest that the internal interconnectedness of an economic system plays a significant role in its overall stability. In the context of regions, this means that regions with diverse economic sectors tend to experience more consistent growth rates than those with fewer economic sectors, highlighting the importance of the interconnectedness between different sectors in determining economic stability.

We then explore a similar pattern for CO$_2$ emissions growth rates as illustrated in Fig.~\ref{fig:sd_size}B. We found that the volatility of CO$_2$ emissions growth rates is inversely proportional to GDP size, with a scaling parameter of $\beta = -0.14$ in the pre-ETS period and $\beta = -0.06$ in the ETS-3 period. This negative relationship implies that higher-income regions tend to display more stable environmental performance, likely due to more efficient energy use, greater technological advancements, and stricter regulatory frameworks. Conversely, smaller or less developed regions experience higher volatility in CO$_2$ emissions, possibly because they lack advanced technological infrastructure and appropriate regulatory frameworks or depend heavily on a limited number of economic sectors. It is critical to recognise that each phase of economic development can introduce significant fluctuations in emissions: new industries may emerge, and existing ones may expand, with the risk of being powered by less efficient and high-emission technologies. Consequently, these regions may become locked into high-emission pathways without adequate policies and investments in clean technologies.

\subsection*{Modelling CO$_2$ emissions growth rates}
%TODO: I think clearer if we add the model equation here.\\
%TODO: here and throughout the paper, clearly mention emission growth rate and GDP growth rate rather than simply "growth rate" to avoid confusion. 

We model the dynamics of the CO$_2$ growth rates of European regions within a dynamic convergence framework, which aims to assess how regions with different economic sizes converge in terms of their emissions growth rates over time. Let $E_{i,t}$ be the CO$_2$ emissions of region $i$ at time $t$. We define the relative size of emissions as $s_{i,t}=\log(E_{i,t})-\overline{\log(E_{i,t})}$, where the second term is the average of emissions (in log) of the European regions, eliminating any possible common emission trend to all regions. The centred emissions growth rates, $r_{i,t}$, are then computed as changes in this relative size. Our dynamic model for $r_{i,t}$ is structured to reflect the discussed relationships between CO$_2$ growth volatility and economic size, as follows: 
\begin{equation}\label{eq:stochastic_ar}
r_{i,t}= \alpha \; s_{i,t-1} + \phi \; g_{i,t} + e^{\beta y_{i,t-1}}\varepsilon_{i,t}\;;
\end{equation}
where $\alpha$ captures the lagged effect of the relative size of emissions $s_{i,t-1}$; $\phi$ the degree of connection with the centred GDP growth rates $g_{i,t}$; $\beta$ the scaling relationship between growth volatility and GDP size $y_{i,t-1}$; and, $\epsilon_{i,t}$ is the regression residual, which also represents the re-scaled emissions growth rates that are assumed Laplacian-distributed and homoscedastic. The parameter $\alpha$ is key to understanding the convergence patterns: $\alpha<0$ indicates that emissions tend towards a long-term stationary state; $\alpha=0$ implies that the dynamics resemble a random walk; and $\alpha>0$ indicates divergence, suggesting that emissions are deviating further from the mean over time. See the methods sections for more details on the definitions and the estimation method. 

We performed estimations of Eq.~\eqref{eq:stochastic_ar} for various periods: full sample, pre-ETS, ETS-1, ETS-2, and ETS-3, with the results presented in Table~\ref{tb:model_periods}. Our econometric analysis highlights significant findings across different regulatory periods. The parameter $\alpha$ consistently indicates a negative relationship across the full sample and during various phases of the EU ETS, suggesting a general trend towards a long-term stationary state in emissions growth rates. However, $\alpha$ was not statistically significant from zero during the ETS-1 phase, indicating an unclear convergence pattern. This lack of significance suggests potential instability or a failure to achieve convergence, which could reflect high uncertainty and challenges in regulatory effectiveness.
%*-*-*-*-*-*-*-*-*-*-*-*-*-*-
\begin{table}[h!]
\begin{center}
\scriptsize 
\caption{Least absolute deviations estimation of the dynamic convergence model for CO$_2$ emissions growth rates across different periods, as defined by Eq.~\eqref{eq:stochastic_ar}.}
\label{tb:model_periods}\vspace{3pt}
\resizebox{0.7\textwidth}{!}{\renewcommand{\arraystretch}{1.1}
\begin{tabular}{l cc ccc } 
\toprule 
 & Full sample & pre-ETS & ETS-1 & ETS-2 & ETS-3 \\
 & (1990-2022) & (1990-2004) & (2005-2007) & (2008-2012) & (2013-2020) \\
\midrule
$\alpha$ & -0.004*** & -0.007*** & -0.001 & -0.001** & -0.005*** \\
 & (0.000) & (0.000) & (0.001) & (0.000) & (0.001) \\
$\phi$ & 0.266*** & 0.189*** & 0.316*** & 0.447*** & 0.364*** \\
 & (0.015) & (0.012) & (0.030) & (0.028) & (0.021) \\
$\beta$ & -0.085*** & -0.118*** & 0.011 & -0.069* & -0.085*** \\
 & (0.011) & (0.022) & (0.048) & (0.038) & (0.029) \\
\bottomrule
\multicolumn{6}{p{9cm}}{\textit{Notes:} Standard errors are in parentheses. Significance level: $^{*}$p$<$0.1; $^{**}$p$<$0.05; $^{***}$p$<$0.01.}\\
\end{tabular}}
\end{center}
\end{table}
%*-*-*-*-*-*-*-*-*-*-*-*-*-*-

The parameter $\phi$ estimations are positive and statistically significant across all periods, indicating a strong link between GDP and CO$_2$ emissions growth rates. The magnitude of $\phi$ is notably higher during ETS-2 and ETS-3, suggesting an increasing influence of GDP growth on emissions as the regulatory framework evolved. This evidence mirrors the critical relationship between economic activity and environmental impact, highlighting the increasing interdependencies between economic growth and emissions.

Similarly, the parameter $\beta$ exhibits a negative significant effect in most estimations, except during the ETS-1 period, reaffirming the robust negative scaling relationship between emissions growth rates volatility and GDP sizes. This parameter's variations suggest changes in the organisation of the economic system, which we can conceptualise as a network of growing and interacting units \cite{lee1998universal, stanley1996scaling, sutton2002variance, bottazzi2006explaining, riccaboni2008size, castaldi2009patterns, campi2020volatility}. Consequently, changes in $\beta$ could be attributed to policy effects, particularly in how carbon market regulations are being adjusted. Nevertheless, one cannot discount the relevance of other known external factors, such as evolving environmental conditions, which may also impact emissions over time.

\subsection*{Dynamic estimation results}

\begin{figure}[h!]
\centering
\includegraphics[width=0.8\linewidth]{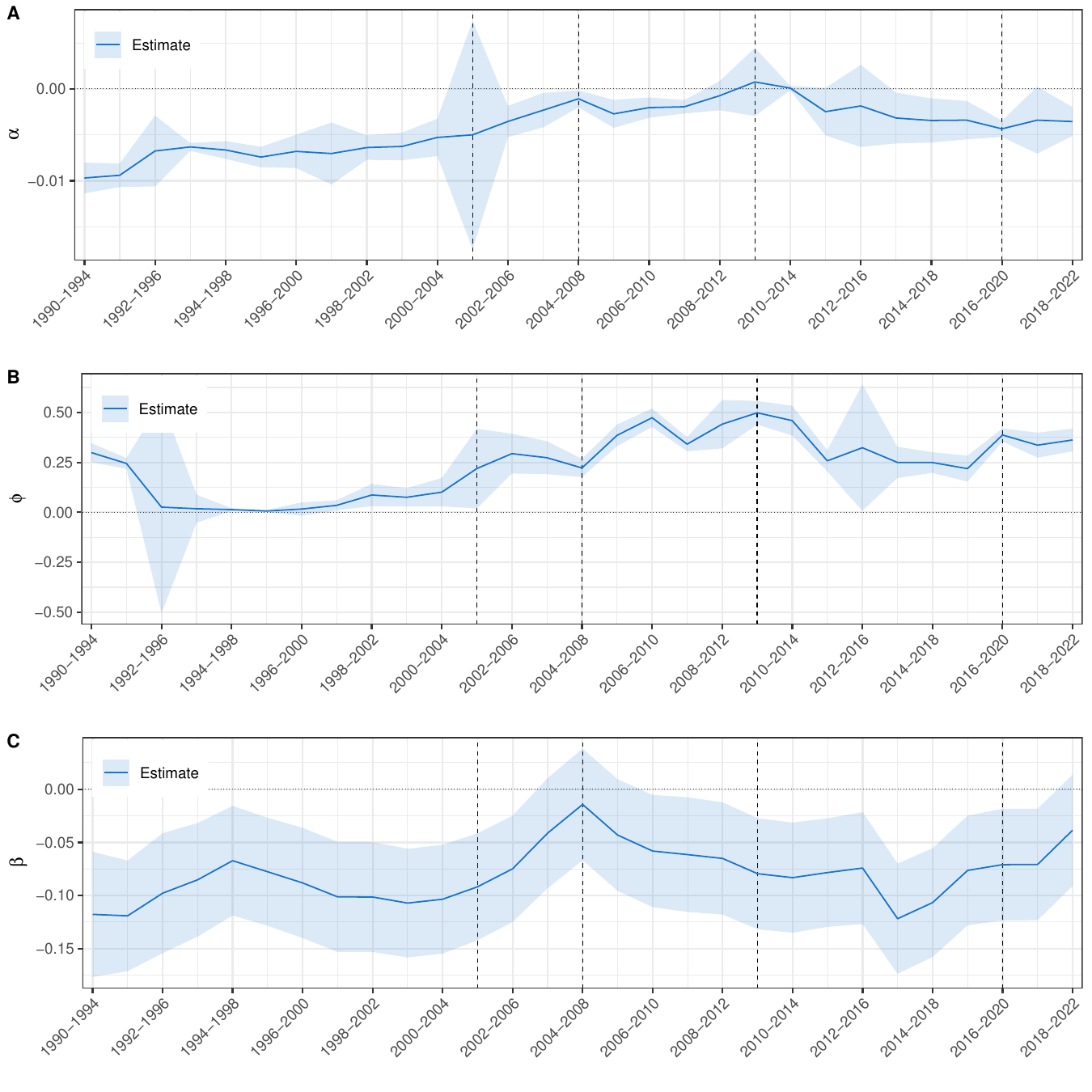}
\caption{Estimations of model parameters across EU ETS phases using a 5-year moving window. Panel A shows parameter $\alpha$, panel B shows parameter $\phi$, and panel C shows parameter $\beta$. Vertical dashed lines indicate the years of implementation of the four different EU ETS phases.}
\label{fig:estimations_mw}
\end{figure}

We estimated the model using a moving window approach. Throughout the analysis, the parameter $\alpha$ generally exhibited negative values, suggesting a trend towards convergence, as shown in Fig.~\ref{fig:estimations_mw}A. However, during certain periods, $\alpha$ was not statistically different from zero, suggesting a scenario akin to a random walk where emissions fluctuated unpredictably without trending towards a long-term equilibrium. Interestingly, although a rising trend of the estimated parameter across the moving windows was observed, the most robust convergence pattern -- indicated by the lowest values -- occurred in the pre-ETS period. Notably, after the initiation of the ETS in 2005, there was a distinct rise to higher values, even positive, in the 2009-2013 moving window. This trend might not be favourable as it could indicate a lack of consistent momentum towards emission reduction goals, reflecting instabilities and inefficiencies of the ETS framework. However, during the third phase of the EU ETS, there was a return to negative values, indicating improved emissions growth rates convergence and reduced uncertainty.
%----------------------------------------
%----------------------------------------

The parameter $\phi$, as depicted in Fig.~\ref{fig:estimations_mw}B, remained consistently positive, reinforcing the strong linkage between economic activities and emissions throughout the ETS periods. Moreover, we observed distinctive changes in the estimated values of this parameter across different ETS phases. Particularly following the initiation of the ETS, $\phi$ assumed significantly higher values. This trend shows that while economic growth has consistently been a significant driver of emissions, the marked changes observed since 2005 suggest the potential effects of the policy interventions, resulting in emission shocks being strongly associated with economic activity. 

The parameter $\beta$, in Fig.~\ref{fig:estimations_mw}C, consistently showed significant negative values across most time window estimations, indicating that larger economies generally exhibit more stable emissions. Although the error bands for this parameter are quite wide -- limiting the possibility of observing significant changes over different periods -- it is noteworthy that a shift towards $\beta=0$ was observed following the implementation of the EU ETS in 2005. Subsequently, $\beta$ returned to levels similar to those seen in the pre-EU ETS period, around $\beta=-0.075$. This fluctuation suggests that while the EU ETS may have initially disrupted emissions stability, the long-term effects align more closely with historical trends. 

Finally, we complete our analysis of the dynamic convergence model by examining the distribution of estimation residuals. These residuals represent the rescaled emissions growth rates and, therefore, capture the common dynamics of emission growth rates across all regions. This analysis is crucial for understanding how the volatility of emissions growth rates evolves collectively beyond individual regional variations.

\begin{figure}[h!]
\centering
\includegraphics[width=0.8\linewidth]{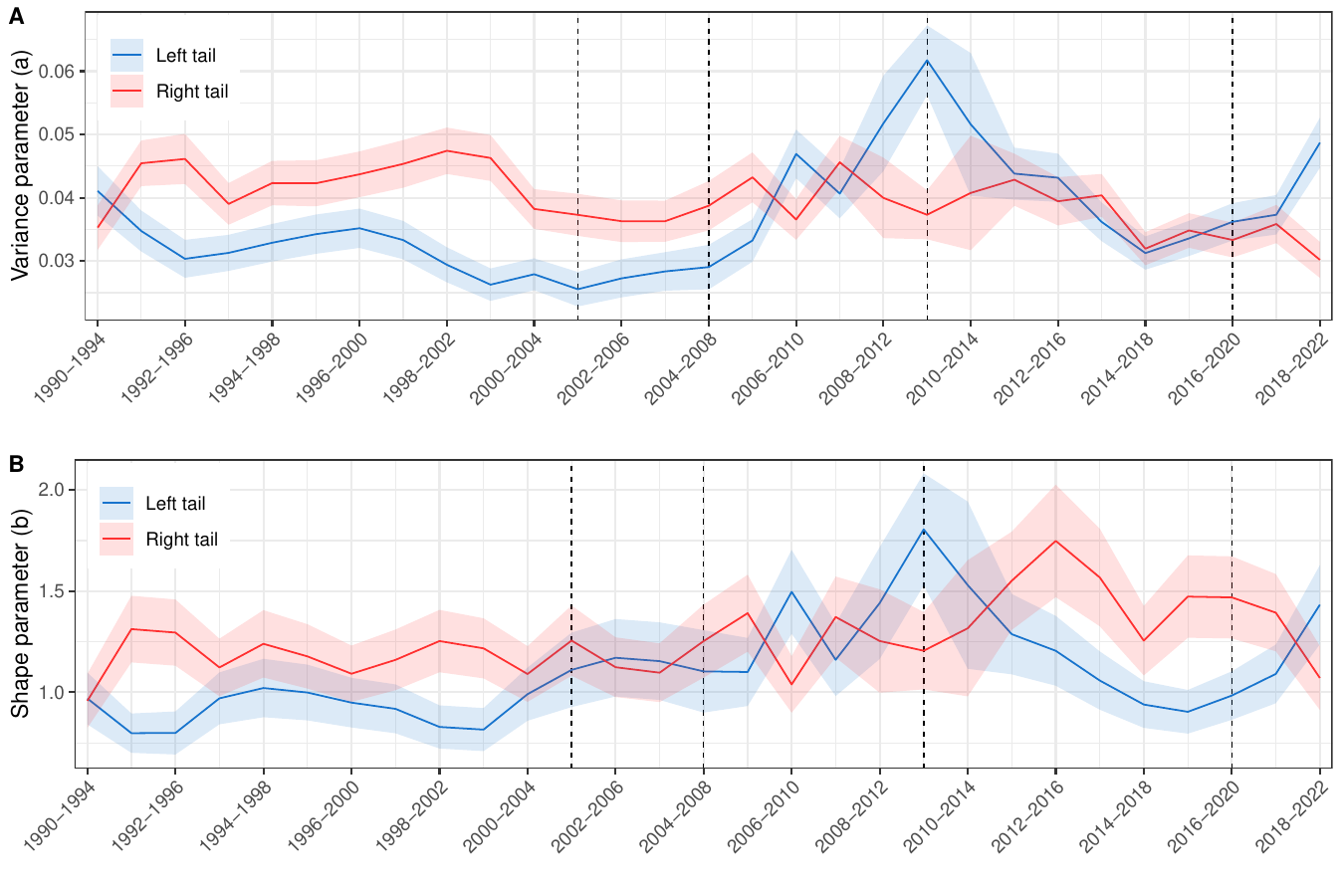}
\caption{Estimations of the AEP distribution parameters of the rescaled emissions growth rates across EU ETS phases using a 5-year moving window. Panel A shows scale parameters $a_{\{l,r\}}$, panel B shows shape parameters $b_{\{l,r\}}$. Vertical dashed lines indicate the years of implementation of the four different EU ETS phases.}
\label{fig:pdf_mw}
\end{figure}

Figure~\ref{fig:pdf_mw}A illustrates the estimated scale parameters $a_l$ and $a_r$, displaying distinct trends. The scale parameter for the right tail, $a_r$, tends to decrease moderately, fluctuating within a narrower range than that of the left tail, $a_l$. This suggests that the left side of the distribution exhibits greater variability over time compared to the right side. Notably, these changes become particularly pronounced from the beginning of the ETS-2 phase onwards. In fact, before the ETS-2 phase, the rescaled distribution is relatively stable and significantly asymmetric, with $a_r > a_l$, persisting throughout almost the entire pre-ETS period and during ETS-1. This latter period, ETS-1, being just the period with lower stringency initially imposed by the EU ETS, corresponds to the three-year pilot phase designed to prepare the EU to meet its Kyoto targets of reducing emissions by at least 5\% between 2008 and 2012 compared to 1990 levels.
%----------------------------------------
%----------------------------------------

Similarly, the shape parameters $b_l$ and $b_r$ exhibit noticeable changes during the EU ETS phases, as depicted in Fig.~~\ref{fig:pdf_mw}B. During the pre-ETS period, there was also an asymmetry in the tail shapes, with $b_r > b_l$. However, the gap between these estimates narrows during ETS-1 and shows less stability following the implementation of ETS-2. Notably, the parameter $b_l$ shifts from a relatively high value of around 1.75 in the 2009-2013 time window to a value close to 0.80 in the 2015-2019 time window. A similar variation occurs for $b_r$ from the 2012-2016 to the 2018-2022 time windows. These transitions from lighter to heavier tails suggest an increased likelihood of extreme emission rates during the corresponding periods.

Towards the end of ETS-3, the distribution appears to become asymmetrical again in the shape of its tails, with $b_r>b_l$ and $b_r>1$. This evidence, coupled with the fact that the parameter $a_r$ has been slowly reducing and is statistically similar to $a_l$, is somehow discouraging in terms of emissions reduction. This implies that while the overall emission levels have been decreasing, there is still a greater probability mass of events above the distributional mode than those below it. This asymmetry in the shape parameters and symmetry in the scale parameters suggest that, near the distribution mode, higher probability mass is associated with positive emission rates than negative. Therefore, albeit in a descriptive manner, this could be interpreted as greater certainty in positive emission shocks, which may not be undesirable but raises significant concerns about the clear progress towards emissions reduction.

Is there a place for policy interventions to change the distribution's tails so that volatility decreases? Given the strong linkage between emissions and economic activity and the significant positive correlation between GDP and emissions growth rates, it seems unlikely that the distribution would exhibit light tails, as fat tails characterise several economic phenomena, including GDP and firm sales growth rates distributions. However, the emerging asymmetry and the excess of volatility are noteworthy, which might indicate that the macroeconomic incentives could be having unintended effects on firm-level behaviour.

\section*{Discussion}

In this paper, we analysed changes in the volatility of CO$_2$ emissions growth rates throughout the various phases of the EU ETS. The EU has significantly enhanced its administrative capacity to meet emissions reduction targets, an aspect that was virtually non-existent in the 1990s. This development has involved integrating lessons learned from each phase of the ETS, leading to substantial institutional changes at the levels of the EU Commission, local governments, and within companies themselves. These transformations have collectively influenced the dynamics of emissions.

Theoretically, policy changes can alter market expectations and business strategies in the short term, creating a more volatile and less predictable environment until the stakeholders adapt to the new conditions. This volatility is exacerbated when political decisions directly impact the operating costs and long-term investment strategies of companies. Such is the case with the EU ETS, where regulatory adjustments are primarily designed to ensure compliance and reduce GHG emissions. Yet, the EU ETS simultaneously aim to promote cost-effective emissions reductions, provide incentives to low-carbon investments, and generate revenues for climate action. Those objectives influence corporate strategies and market behaviour, underscoring the complex impacts on emissions. 

This study provides a detailed analysis of the evolution of volatility in CO$_2$ emissions growth rates without establishing explicit causal relationships between policy changes and observed emissions dynamics. However, by discussing our results within the historical context of EU ETS policy developments, we provide a comprehensive overview of how such policies might have influenced emissions dynamics over time.

Our estimations indicate that after the implementation of the EU ETS, the emissions dynamics were significantly influenced, with more pronounced changes observable from the second phase onwards. There are several reasons to think that the first phase (ETS-1) had a mild impact on volatility, yet it introduced significant uncertainty in the emissions dynamics. The estimations reveal that during ETS-1, there was no evident convergence pattern in the dynamics of emissions growth rates, and the impact of GDP growth on emissions became more pronounced compared to the pre-ETS period.

This lack of convergence during ETS-1 highlights the challenges of the initial phase, which was intended as a learning phase to familiarise companies with emissions trading without imposing high initial costs. However, it yielded little direct commitment from companies to reduce emissions \cite{neuhoff2010implications} and was notably compromised by political influence and business lobbies. These groups favoured the free allocation of emissions rights and prioritised specific interests over broader environmental objectives \cite{sato2022allocation}, further diluting the phase's effectiveness.

Likewise, before the implementation of the EU ETS, the EU had collected little information on CO$_2$ emissions at the facility level. This lack of data led to significant information asymmetries between national regulators and polluters \cite{schmalensee2017lessons}, enabling companies to mitigate the effects of imposed emissions restrictions, at least in the short run \cite{grubb2006allocation}. Consequently, this phase restricted the ``banking'' of unused European Union Allowances (EUAs) for use in subsequent phases. 

The ETS-2 phase marked an improvement in the use of emissions data at the installation level, which is crucial for guiding the National Allocation Plans. This phase, aligning with the first commitment period of the Kyoto Protocol, saw the European Commission increasing regulatory rigidity, which led to significant tensions among stakeholders \cite{betz2010eu}. Despite historically low carbon prices during this period, the new bureaucratic challenges associated with obtaining restricted EUAs likely influenced emissions dynamics \cite{sanin2015}. In our estimations, the convergence pattern during ETS-2 was relatively weak compared to the estimation of other periods, though still significant, with the impact of GDP growth rates becoming more prominent. 

From a bureaucratic perspective, ETS-3 introduced the most significant changes. Key modifications included the centralisation of the allocation process under the supervision of the European Commission, the adoption of auctioning as the primary method for distributing allowances, the implementation of the Carbon Leakage Decision Mechanism, and the establishment of the Market Stability Reserve. These changes imposed stricter constraints on companies, yet they also established clearer rules and reduced policy uncertainty compared to previous phases \cite{sato2022allocation}. Our estimations indicate that during ETS-3, the pattern of convergence dynamics was strengthened. Moreover, the impact of GDP growth rates continued to remain high, similar to that observed during ETS-2.

Our study also provides further insights into the volatility dynamics by tracking the evolution of the probability distribution of the rescaled emissions growth rates and the changes in the scaling $\beta$ parameter. The rescaled emission rates reveal that volatility dynamics are driven by significant changes in both tails of the distribution. Therefore, both extreme decreases and increases in emission rates contribute to the overall volatility. 

On the other hand, we observed that the $\beta$ parameter starts increasing from the implementation of the EU ETS and approaches zero during the transition from ETS-1 to ETS-2. This empirical evidence suggests no significant differences in volatility among regions of different economic sizes. This implies that, while decarbonising their economies, smaller economic regions are becoming less volatile or, conversely, larger economic regions are becoming more volatile. The evidence tends to support the latter scenario. In fact, larger regions, typically in higher-income countries, may have demonstrated greater adherence to ETS policy initiatives. This observation is consistent with the notion that regions hosting larger companies faced more stringent demands to reduce emissions swiftly. This connection is further supported by environmental indicators such as CO$_2$ emissions, energy use, water extraction, and waste production, which tend to follow a power law relationship with company size \cite{mastrandrea2024assessments}. Consequently, large companies, which significantly influence GDP growth volatility \cite{gabaix2011granular}, also play a pivotal role in shaping the observed emissions dynamics.

Nevertheless, while higher volatility associated with larger regions could explain the empirical evidence, compositional growth models could also shed light on the effects of policy on how volatility is distributed across stakeholders. These models suggest that increasing interdependencies within a system composed of different growing entities can affect output growth rates' volatility. Specifically, volatility tends to decrease when parts of the system are highly interdependent, indicating fewer differences in volatility across the system \cite{lee1998universal, stanley1996scaling, sutton2002variance, bottazzi2006explaining, riccaboni2008size, castaldi2009patterns, campi2020volatility}. Thus, the results of our analysis imply that during the ETS-1 and ETS-2 periods, the regulations led to greater interdependence among companies. This likely emerged because the new mechanism, which introduced a cap on emissions, inherently linked companies' decisions, thereby increasing their interdependence.

The dynamics of the $\beta$ parameter moving towards significantly negative levels between ETS-2 and ETS-3 suggests a return to the pre-ETS ``normality'', which can be associated with the typical growth pattern of heterogeneous entities. This shift might reflect the establishment of more consistent and transparent rules in the allocation of emission allowances, meaning that companies have effectively incorporated learning from previous phases and adapted to the regulatory environment.

As a final remark, the interplay between economic policy, competition, and technological change is critical in shaping companies' growth trajectories and regional development. The core challenge is to decouple economic growth from emissions within the business cycle's dynamism, formulating strategies that promote sustainable development without compromising socioeconomic progress.

\section*{Methods}

\subsection*{Data}

We collect emission data from the Emissions Database for Global Atmospheric Research (EDGAR), managed by the Joint Research Centre of the European Commission. EDGAR provides quantitative estimates of global GHG emissions using a methodology that aligns with the Intergovernmental Panel on Climate Change (IPCC) guidelines \cite{EDGAR_report}. Specifically, we use data on aggregated GHG emissions at the European regional NUTS 2 level of aggregation. Additionally, we use demographic and economic data from ARDECO (the Annual Regional Database of the European Commission's Directorate General for Regional and Urban Policy), which offers annual regional data maintained and updated by the Joint Research Centre. Our dataset spans from 1990 to 2022, covering 242 regions.

\subsection*{Size and growth rates definitions}
To model the growth rates of CO$_2$ emissions, we employ a statistical approach that uses variables expressed in relative sizes. Given that GHG emissions are proportional to economic activity, analysing regional emissions outcomes without adjusting for economic size could introduce biases when comparing regions of differing economic scales. Therefore, using relative sizes offers several advantages. It enables more accurate comparisons across regions by neutralising variations in economic size, and it helps mitigate the potential impact of outliers in the statistical model, which could otherwise lead to inconsistent findings.

Therefore, for a region $i$ at year $t$, we define the relative emission size as
\begin{equation}\label{eq:size}
  s_{i,t}=\ln(E_{i,t})-\frac{1}{N}\sum_i{\ln(E_{i,t})} \;;
\end{equation}
where $E_{i,t}$ are the total emission and $N$ is the number of regions. Notice that this definition eliminates any possible trend common to all regions. Based on this, the corresponding emissions growth rates are defined as changes in the relative size as
\begin{equation}\label{eq:gr_r}
  r_{i,t}=s_{i,t}-s_{i,t-1} \; .
\end{equation}
Note that since the size measure in Eq.~\eqref{eq:size} is centred around zero, the growth rates are, by construction, centred around zero for any $t$.

Similarly, we define GDP in terms of its relative size as
\begin{equation}\label{eq:size_gdp}
  y_{i,t}=\ln(GDP_{i,t})-\frac{1}{N}\sum_i{\ln(GDP_{i,t})} \;.
\end{equation}
Consequently, the GDP growth rates are defined as changes in its relative size as 
\begin{equation}\label{eq:gr_gdp}
  g_{i,t}=y_{i,t}-y_{i,t-1}.
\end{equation}

\subsection*{Parametric distribution of growth rates}
% \cite{pena2022log, pena2023gibrat, ahundjanov2019gibrat} REVISAR LA DISCUSION QUE HACEN SOBRE GIBRAT??.
We adopt an asymmetric exponential power density function to analyse the probability density functions of growth rates \cite{bottazzi2011new}, specifically
\begin{equation}\label{eq:subb}
  f(x;a_l,a_r,b_l,b_r,m)=
    \frac{1}{A}  
    e^{-\left(\frac{\theta(m-x)}{b_l}\left|\frac{x-m}{a_l}\right|^{b_l} + 
    \frac{\theta(x-m)}{b_r}\left|\frac{x-m}{a_r}\right|^{b_r}
    \right)}
%  \begin{cases}
%    \frac{1}{A}e^{-\frac{1}{b_l}|\frac{x-m}{a_l}|^{b_l}} & x<m  \\
%    \frac{1}{A}e^{-\frac{1}{b_r}|\frac{x-m}{a_r}|^{b_r}} & x>m
%  \end{cases} \; ;
\end{equation}
with
\begin{displaymath}
  A=a_lb_l^{1/b_l}\Gamma(1+1/b_l) + a_rb_r^{1/b_r}\Gamma(1+1/b_r) \; ;
\end{displaymath} 
where $a_{\{l,r\}}>0$, $b_{\{l,r\}}>0$, $\theta(\cdot)$ is the heaviside function, and $\Gamma(\cdot)$ is the gamma function. The parameter $m$ is the position of the mode, the scale parameters $a_{\{l,r\}}$ characterise the statistical dispersion of the left and the right tails, while the shape parameters $b_{\{l,r\}}$ describe their asymptotic behaviour. The symmetric version of this density is recovered when the left and right parameters are equal. We estimate Eq.~\eqref{eq:subb} via maximum likelihood estimation. 

These AEP densities can lead to more traditional density functions. For instance, when $a_l=a_r$ and $b_l=b_r=2$, the growth rates follow a normal distribution. Conversely, if $b_l=b_r=1$, then the growth rates fit to a Laplace distribution. Referencing these two density functions is useful for understanding the dynamics of growth rates. Theoretically, the probability of encountering extreme events, such as sudden slumps, is much lower if growth rates are derived from a normal distribution than from a Laplacian distribution. Specifically, the lower the $b_{\{l,r\}}$ parameters, the higher the likelihood of finding extreme events. Therefore, looking at the tails, as indicated by the parameters $a_{\{l,r\}}$ and $b_{\{l,r\}}$) around the mode ($m$), provides insightful information for analysing the distribution of growth rates. 

\subsection*{Statistical model}
We model the growth rates of CO$_2$ emissions within a dynamic convergence framework as detailed in Eq.~\eqref{eq:stochastic_ar}, along with empirical evidence of fat-tails in the distribution of emissions growth rates. The model tests if regions with higher emissions levels tend to reduce them over time, thus diminishing regional disparities. Additionally, it considers the pro-cyclical nature of emissions relative to GDP growth fluctuations, testing the hypothesis that output shocks have immediate impacts on emission rates. This procyclicality accounts for the dynamic interaction between economic activity and environmental impact, which has relevant policy implications in promoting sustainable growth.

In addition, the model considers the scaling relationship between the dispersion of emissions growth rates and economic size. This relationship suggests that $\ln(\sigma_g) \sim \beta y$, where $\sigma_g$ is the standard deviation of GDP growth rates. Given the close link between economic output and emissions dynamics, we propose a similar scaling relationship for emissions
\begin{equation}
 \label{eq:beta}
 \ln(\sigma) \sim \beta y\;,
\end{equation}
where $\sigma$ represents the standard deviation of emissions growth rates. From a statistical standpoint, this scaling relationship indicates the presence of heteroscedasticity, which violates the homoscedasticity assumption of classical econometric models that rely on ordinary least squares (OLS). 

To estimate Eq.~\eqref{eq:stochastic_ar}, we adopt a nonlinear least absolute deviations approach, formulated as follows:
\begin{equation}
\label{eq:MAD}
\text{argmin}_{\{ \alpha, \phi, \beta\}} \sum_{i}\sum_{t}
\left|
\frac{r_{i,t} - \alpha \; s_{i,t-1} - \phi \; g_{i,t} }{e^{\beta y_{i,t-1}}}
\right| \;.
\end{equation}
The least absolute deviations approach is preferred over OLS estimators because it is better suited for data with potential heavy tails, offering more robust and reliable results.

\section*{Acknowledgements}
The authors acknowledge the financial support of the DECIPHER (Decision-making framework and processes for holistic evaluation of environmental and climate policies) project, funded by the Horizon Europe: Research and Innovation programme under grant agreement No. 101056898.

%\clearpage
%\newpage
%\small
%\bibliographystyle{unsrt}
\bibliographystyle{ieeetr}
\bibliography{biblio_fat}

%\section*{Author contributions statement}
%Both authors conceived the study and conducted the methodology. M.D. performed the analysis and analysed the results. Both authors contributed to reviewing the manuscript.

\clearpage
\newpage
\appendix
\section*{Supplementary information}

\setcounter{table}{0}\renewcommand{\thetable}{SI.\arabic{table}}
\setcounter{figure}{0}\renewcommand{\thefigure}{SI.\arabic{figure}}
\setcounter{equation}{0}\renewcommand{\theequation}{SI.\arabic{equation}}

\subsection*{Evolution of GDP and emissions growth rates}

\begin{figure}[h!]
\centering
\includegraphics[width=\linewidth]{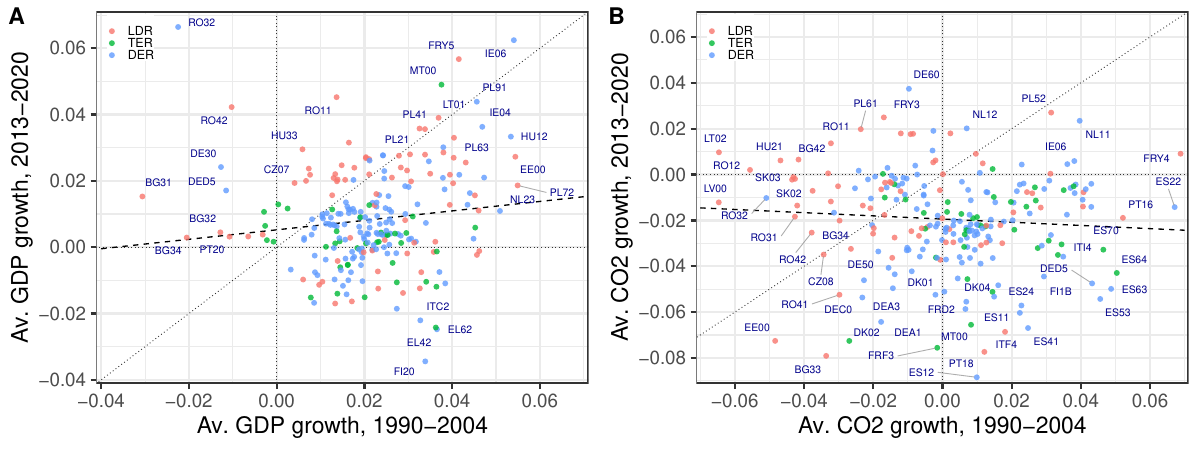}
\caption{Average GDP and emissions growth rates scatter plots between 1990-2004 (pre EU ETS) and 2013-2020 (EU ETS-3) periods. Panel A shows GDP growth rates. Panel B shows CO$_2$ emissions growth rates. In both panels, regions are classified by development level: less developed regions (LDR), transition regions (TER), and developed regions (DER). A dashed line represents the average trend.}\label{fig:av_plots}
\end{figure}

Since the 1990s, economic growth and greenhouse gas emissions in European regions have changed significantly. Figure~\ref{fig:av_plots} compares the average growth of GDP and CO$_2$ emissions for the periods 1990--2005, the pre-EU ETS period, and 2013-2020, the EU ETS-3 period. In these plots, we have used colours to distinguish regions with different income levels: less developed region (LDR), transition region (TER), and developed region (DER).\footnote{We use the classifications reported in annexes in: \url{https://eur-lex.europa.eu/legal-content/EN/TXT/?uri=CELEX:32014D0099}}
%----------------------------------------
%----------------------------------------

Figure~\ref{fig:av_plots}A shows that most regions experienced a slowdown in GDP growth between the pre-EU ETS and ETS-3 periods. However, while dispersed, the overall trend remains positive. At first glance, as expected, high GDP growth is associated with less developed regions; many LDRs are above the trend line, and some even exhibit higher growth rates (above the 45-degree line). This contrasts with the more developed regions, for both TER and DER, which are less dispersed in the scatter diagram.

Regarding the CO$_2$ emissions growth rates, see Fig.~\ref{fig:av_plots}B, the pattern is also dispersed. The overall trend is negative, contrasting with the trend observed for GDP growth rates, with TERs and DERs now appearing less concentrated. However, while a few regions are located in the positive quadrant ($x>0$ and $y>0$), indicating a generalised tendency to decelerate emission rates, it is noteworthy that a large number of regions are above the 45-degree line. These regions increased their emissions growth rates between the two periods. In this group, which shows comparatively poorer improvements in reducing emission rates, there are many LDCs as well as TERs and DER.

\end{document}